# Tales of 34 iPhone Users:
## How they change and why they are different

Ahmad Rahmati[1], Clayton Shepard[1], Chad Tossell[2], Mian Dong[1], Zhen Wang[1], Lin Zhong[1], Philip Kortum[2]
[1] Department of Electrical & Computer Engineering,  [2] Department of Psychology
Technical Report *TR-2011-0624*,  Rice University, Houston, TX

## Abstract

We present results from a longitudinal study of 34 iPhone 3GS users, called LiveLab. LiveLab collected unprecedented usage data through an in-device, programmable logger and several structured interviews with the participants throughout the study. We have four objectives in writing this paper: (*i*) share the findings with the research community; (*ii*) provide insights guiding the design of smartphone systems and applications; (*iii*) demonstrate the power of prudently designed longitudinal field studies and the power of advanced research methods; and (*iv*) raise important questions that the research community can help answer in a collaborative, multidisciplinary manner.

We show how the smartphone usage changes over the year and why the users are different (and similar) in their usage. In particular, our findings highlight application and web usage dynamics, the influence of socioeconomic status (SES) on usage, and the shortcomings of iPhone 3GS and its ecosystem. We further show that distinct classes of usage patterns exist, and these classes are best served by different phone designs, instead of the one-size-fits-all phone Apple provides. Our findings are significant not only for understanding smartphone users but also in guiding device and application development and optimizations. While we present novel results that can only be produced by a study of this nature, we also raise new research questions to be investigated by the mobile research community.

## 1. Introduction

We present findings from an unprecedented, longitudinal study of 34 iPhone 3GS users, 24 for 12 months and 10 for 6 months, called LiveLab [12]. The study leveraged an in-device, programmable, continuously running logger that collects device usage. The logs are further enhanced by regular interviews with the participants. To the best of our knowledge, LiveLab is unique in the following important ways.

First, LiveLab is the first publicly reported study of smartphone users with in-device logging over six months. In contrast, prior work lasted at most a few months. The yearlong study allows us to study the adoption and long-term evolution of user behavior that has been previously impossible.

Second, unlike prior work that has very limited information of or interaction with the participants [1, 2], our study features carefully selected participants with controlled demographics and carefully designed interaction with them over the course of the study. Instead of trying to represent a broad demography of smartphone users, we chose to focus on a very specific user population, college students of similar age, but with different socioeconomic backgrounds. This strict selection allows us to gain deep insight in to the behaviour of this population, as well as discover the unadulterated influence of socioeconomic status on usage. Our unique access to the participants further allows us to gain otherwise impossible insights into the data collected by the in-device logger.

Third, LiveLab is the first publicly reported study of iPhone users with in-device usage logging. Prior work has studied usage of Android and Windows Mobile based smartphones with in-device logging. We chose the iPhone as it represents the cutting edge of smartphone design for usability, accounting for over a third of the US mobile internet traffic as of April 2010 [3]. Additionally, iPhone users have access to the largest number of third-party applications, with over 300,000 officially released apps as of October, 2010. Our study is the first to provide a comprehensive picture of how iPhone users employ their devices in real environments and leverage the Apple App Store.

Sitting atop a goldmine of data, we refuse to simply present the usage statistics in this paper. Instead, we selectively provide insightful findings that are only possible with the three unique features highlighted above. Our findings include not only the large diversity in participants' application usage, but also the long-term variation, seasonal trends, as well as intra-user and inter-user similarities. We find that most of the usage by a participant converges on a small number of applications and websites. Our findings indicate that smartphone web usage is more of an extension to the users' PC based web access and users' disappointment with the web browsing experience on smartphones decreases their usage. While our core 24 participants all attend the same small private college and live in similar dorms, a seemingly homogenizing environment, we show that their socioeconomic status still has a significant impact on several aspects of usage likely because users with different SES brackets have different needs and preferences within their particular contexts. This led to dissimilar usage patterns. We show that while smartphone users are known to be diverse [4], at the same time they can be quite similar. Our clustering analysis reveals four distinct groups of usage patterns, consisting of two large clusters and two outliers, even for the 34 college students. The distinct clusters high-light the intra-group similarities and suggest the need for multiple distinct hardware and software customizations for smartphones.

Our findings have strong implications for not only understanding smartphone users, but also for device and application design, optimization, and evaluation. We show the importance of long-term user studies with carefully selected participants, and highlight the benefits of having interviews with at least a fraction of participants. We demonstrate the importance of a try-before-you-buy App Store, while showing that having web based versions of applications, whenever possible, facilitates users to try them out. We show the feasibility and limitations of smartphones for IT access, including for cost effective IT access for underserved communities. In particular, our results strongly suggest smartphone users could benefit from a better web browsing experience. Last but not least, we debunk the myth of a one-size-fits-all phone, and show that even among our limited set of participants, there are distinctively different usage patterns that would benefit from phones with different hardware / software configurations.

In addition to leveraging the unique strengths of this study to produce novel results, we hope to shed light on new and important questions that remained unanswered regarding mobile device usage, socioeconomic status, and user diversity.

The rest of the paper is organized as follows. We discuss related work on field studies of smartphone usage in Section 2. We describe the field study and our methods for data analysis in Section 3. From Section 4 to 6, we present findings regarding application usage, web usage, and the impact of socioeconomic status, respectively. In Section 7, we employ Ward's clustering method to understand how our 34 participants are different and similar. We offer the design implications of our findings in Section 8 and conclude in Section 9.

## 2. Related Work

Human factors of mobile devices have been an active research area for more than a decade. Most human factors studies employ either lab-based evaluation or a short period of field trials. In the last few years, as smartphones began to be widely adopted, there have been several relative long-term field studies of smartphone usage that are related to ours.

In [5], the authors studied 12 high-school users of Windows Mobile smartphones (HTC Wizard) for four months and a control group of 10 college users for one month. The in-device logger used in the study only recorded the screen status and network conditions, providing very limited information regarding usage and its context. Moreover, the two groups of participants were different in many aspects other than socioeconomic status and, therefore, many observations regarding socioeconomic status impact were merely speculative with only qualitative evidence. In contrast, our study, with a much longer period, a much more powerful logger, and carefully selected participants, provides much deeper insights and conclusive findings regarding the usage and the influence of socioeconomic status.

In [4], the authors studied 33 users of Android smartphones for 7 – 21 weeks. The authors did not have access to the participants for interviews or have demographic information about them beyond several predetermined user types. The data was analyzed mostly for the usage statistics in the form of distributions. The major conclusion made was that smartphone users are very different without providing insights into why. In contrast, our study employed participants selected in a controlled manner, and a much longer period of study (12 months vs. 2 months). The careful selection of participants and the longer period of study enable us to apply novel analysis techniques beyond simple statistics to gain insights into the long-term evolution of smartphone usage and into answering the question why smartphone users are different (and similar). Moreover, with a superset of usage data, we are also able to analyze many new aspects of smartphone usage, including App Store utilization, application usage, and web access.

Multiple research efforts have utilized data collected by a cellular network carrier to study usage location [6], as well as call patterns [7] and statistics [8]. However, usage data collected by network operators is limited in both scope and detail. For example, cellular network carriers are unable to collect data for applications that do not access the network, or when a user is using WiFi. The study reported in [6] focused on the location/mobility of mobile access to websites through the cellular network without knowing the demography of the users. Moreover, it was limited to only the top-level domain names of web access. In contrast, we collected the URL of each web access, which provides key insights into the content of smartphone web access.

The MIT Reality Mining project [9] studied 100 users of Nokia Symbian 60 series phones for one year. The study employed an in-device logger to record information regarding where the user was and how close a user was to another (using Bluetooth). Coupled with self-reported information, the study sought to reveal social relationships between the subjects. In contrast, our study focuses on the usage of the device. Therefore, the data we collected over one year is much more extensive. For example, the compressed size of our data is two orders of magnitude larger than that of the Reality Mining data (10 GB vs. 55MB). Furthermore, the Nokia is a previous generation smartphone and their usage do not generalize to current generation smartphones (e.g., 80% of their device usage was for voice phone calls).

In [10], the authors collected resource usage data from Android smartphones for one month, and analyzed the power usage by various hardware components using a system power model constructed in the lab. Similarly, the authors of [11] observed simple usage statistics of 15500

Table 1: We assigned categories to applications based on the genres reported by the App Store

| Category | Genres | Notes |
|---|---|---|
| Games | Games, Entertainment, Media | Entertainment and media consumption |
| Utilities | Utilities and Productivity | Calculators, alarm clocks, todo lists |
| Reference | Books, Education, and Reference | Information resources |
| News | News, Sports, Travel, Weather | Contemporaneous information resources |
| Commerce | Business, Finance, Lifestyle (shopping) | Shopping or financial apps |
| Social Networking | Social Networking | Facebook, MySpace, Twitter |
| Other | Health, Navigation, Medical, Photography | Only a few (162) applications |

Blackberry users for different periods, on average 29 days, for the purpose of battery management and prediction. The authors of [1] deployed a network testing application through the App Store and the Android Market to a large number of users. When users run the application, it measures the network performance and reports the measurements. These studies focused on the power usage by hardware and network performance characterization, which are complementary to our focus on the usage and the users.

# 3. Field Study and Data Analysis

Our field study, LiveLab, lasted from February 2010 to February 2011 with 24 iPhone 3GS users and from September 2010 to February 2011 with 10 more users. We next provide details regarding the study.

## 3.1 Field Study Participants

Users in the study were young college students (average age: 19.7, deviation: 1.1). The core 24 participants studied for one year were recruited from two distinct socioeconomic (SES) groups from a small private university at a major metropolitan area in the USA. They all lived on campus and in similar dorms. 13 received need-based scholarships and 11 did not. We used this information to separate the former into a low SES group and the latter into a high SES group. There was no significant bias in the participants, including their major, gender, race, PC access, and game console ownership. All had a PC or laptop at their residence, in addition to access to the university's computing labs. 11 of the low SES participants and all high SES participants had a personal laptop.

Approximately six months into the study, we extended the study with 10 students from a community college located in an underserved part of the same metropolitan area. The users from this campus were classified as Very Low SES and only have six months of usage logs per user. Due to the difference in study length and campuses, most of our analysis will focus on the core 24 participants from the private university. The 10 community college participants will only be used in Section 7.

Every participant received a free iPhone for their participation. Additionally, each participant received free service coverage, including 450 voice call minutes per month, unlimited data, and unlimited SMS for the entire time data were logged. We helped all participants port their phone numbers to the iPhones and they were required to use the outfitted iPhones as his or her primary device.

## 3.2 Logger Design and Implementation

The key component of the field study is an in-device, programmable logging software that collects almost all aspects of iPhone usage and context *in situ*. To run the logger in the background continuously, we had to jailbreak the iPhone 3GSs and exploit a setting provided by the iOS that starts the daemon process, as well as restarts it anytime it is killed. The main logger daemon is written as a shell script in bash and utilizes components written in various languages, including C, perl, awk, SQL, and objective C. Furthermore, the logger daemon is able to call built in functions, manage child processes, install and use programs from repositories, run custom programs, and add new features. We have implemented the logger in a modular and robust fashion, thus a new iOS release may break individual components, but the main functionality will not be affected. In order to monitor and update the logger, it is programmed to report data and, if necessary, update itself every day through an encrypted connection, via rsync [13], to a lab server. We employed several methods to limit energy consumption, and our measurements show that the logger consumes on average less than 5% of the phone battery per day.

While the logger records a plethora of context information, for this work we focused on logs regarding application installation, uninstallation, price, genre, and usage, as well as web usage.

### 3.2.1 Assuring Privacy

Collecting data from smartphones in the field naturally incurs privacy issues. We employ the following methods to protect privacy while retaining relevant information for research. First, we leverage one-way hashing to preserve the uniqueness of a data entry without revealing its content. For example, we hash the recorded phone numbers, names, and email addresses. With hashing, we can still construct call statistics without knowing actual phone numbers. Second, we perform information extraction in the device. For example, we extract emoticons from emails and text messages without collecting the raw content. Finally, we structure the research team so that the data analysis and logger development team do not directly interact with the participants, in order to avoid linking data to the actual users. A separate human factors team acts as the interface with our participants but does not deal directly with the logger or access the raw data. This enables us to contact the partici-pants in a privacy sensitive manner, which we have found to be necessary on numerous occasions, e.g., to schedule impromptu interviews with users who exhibit a drastic change in behaviour.

### 3.3 Complementary Interviews

Since our study design allowed us to have access to the participants, we utilized qualitative interviews alongside automated logging of usage. In particular, interviews are necessary to compare the user perception and their usage, and to distinguish usage changes from system glitches. Interviews are also necessary to assess the participants other IT access methods and previous experiences.

### 3.4 Data Analysis and Methodology

In this section, we present our data analysis methodology.

#### 3.4.1 Defining Similarity

To objectively measure similarity in usage, a uniform and general purpose metric for similarity in usage is necessary. We consider usage as a vector, each element being a specific dimension, e.g., an application or a website. The similarity metric must be applicable to different types of usage, such as application and web usage. The similarity metric must also be applicable to different metrics of usage. For example, we apply the similarity metric to the frequency and/or the duration of application and web use. We also intend to separate the magnitude of the usage vector from its *direction*. Otherwise, the magnitude of usage would be dominate the similarity index and mask the comparatively subtle differences in directions of usage.

Many metrics have been proposed in prior literature for defining a similarity index. They are often based on the distance, both 1-norm (block) and 2-norm (Euclidean), or the angle (Cosine Similarity) between the corresponding usage vectors. Other metrics are based on the properties of their corresponding sets. We chose the Cosine Similarity metric since it ignores magnitude and only considers the direction of the vectors. The Cosine Similarity is the Cosine function of the angle between the two usage vectors, and its output is always between 0 and 1. In mathematical terms, since we have

$$A.B = \|A\|\|B\|\cos\theta$$

we can calculate Cosine Similarity ($S_{A,B}$) as:

$$S_{A,B} = \cos\theta = \frac{A.B}{\|A\|\|B\|} = \frac{\sum_{i=1}^{n} A_i.B_i}{\sqrt{\sum_{i=1}^{n} A_i^2} \cdot \sqrt{\sum_{i=1}^{n} B_i^2}}$$

where $A_i$ and $B_i$ is the amount of usage type $i$ for the users $A$ and $B$ respectively.

Note that the same Cosine Similarity metric can also be used to compare a user with the mean or median of a group of users, or to measure the similarity of one person's usage over different time periods. In particular, we extensively use this similarity metric to identify trends and changes in different months. We use one month intervals as they provide a good balance between detailed information regarding trends and having enough data to draw significant conclusions.

#### 3.4.2 Missing Data

Due to the nature of the study, there were short-term lapses in the log files of five users. The lapses lasted from a few days up to over a month, and were caused by a number of reasons. These include bugs in our code, lost, stolen or damaged phones, travel, and phones that were accidentally restored by the users. We substitute data from missing days with the all time average of that user in order to maintain each user's uniqueness and to avoid magnifying the impact of short-term fluctuation in usage. We note that since missing data only happens for short periods and on few users, and considering the fact that we regenerate the missing samples and analyze macro-dynamics, i.e. long-term (e.g. monthly) usage of users, the overall effect of missing data is negligible.

#### 3.4.3 Outliers

We mitigate the effect of outliers by using median instead of average for our analysis. Due to the limited number of participants and extreme diversity, the effect of outliers can be huge. For example, one of our low SES users told us she used her iPhone exclusively to get on web during the summer, to avoid paying for internet at her summer residence. This caused significantly higher web usage because she could not use her PC or laptop to access the internet. As another example, two low SES participants used the iPod application significantly more, and one high SES almost exclusively used the phone for voice calls. While such examples are anecdotal evidence of certain usage patterns, they can polarize averages for all but extremely large data sets. Using the median can mitigate the effect of these outliers, and present us a better understanding of the *typical user*.

## 4. Application Usage and Dynamics

From our logs, we are able to extract the time each application is installed, uninstalled, and used, as well as application details such as price and genre. We assign categories to applications based on the 20 genres reported by the App Store, as shown in Table 1. In this section, we present findings regarding the adoption and usage of applications, both built-in and from Apple App Store. Our analysis highlights the dynamics of application usage, both in terms of adoption, usage, and duration

### 4.1 Application Purchase and Adoption

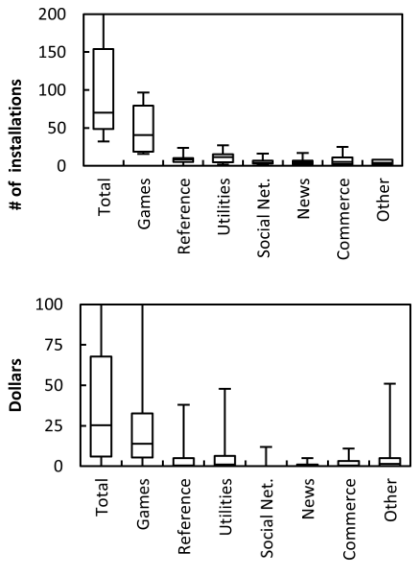
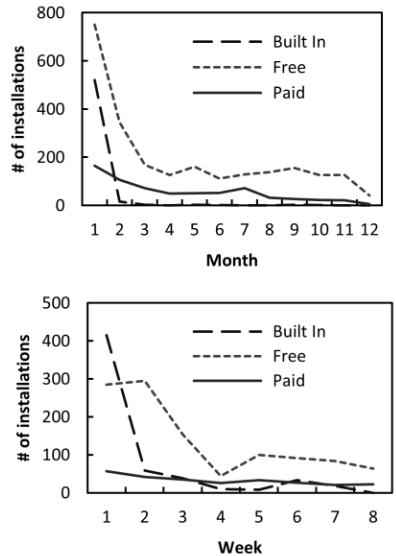

Figure 1: Installed applications broken down by category, in terms of number (top) and price (bottom). Boxes: 2nd / 3rd quartiles. Whiskers: maximum / minimum. Horizontal lines: median

Figure 2: Application adoption for paid, free, and built in applications, monthly for 12 months (top) and zoomed-in weekly for the first 8 weeks (bottom)

Figure 3: Application lifespan by category (top, some categories redacted for clarity) and by price (bottom)

Our 24 participants installed over 3400 applications over the course of the study, of which over 2000 were unique. Our participants also purchased almost 750 applications, of which 500 were unique, from the Apple App Store, spending over $1300. As expected, there was a wide variation between users and application categories. Our users spent a median of $25 on 14 applications, as shown in Figure 1, and all but two users purchased at least one application.

The first two months see a huge number of applications being adopted, highlighting the problem with giving out smartphones and studying the users for only a few months. We define adoption as the time when the user installs a new application from the App Store, or for built-in applications, the first time the user runs it. While our users exhaust almost all built-in applications in the first two weeks, they continue to adopt new applications throughout the study. Figure 2 shows the total number of adopted applications during the study, broken in to built-in, free and paid applications. The ratio of paid to free applications stays relatively constant over time, at around 20%.

The most popular application category was games, accounting for over 50% of application installs and over 50% of money spent, and approximately 5% of application usage. In contrast, social networking applications, mostly being free, only accounted for less than 2% of money spent, but accounted for 8% of application usage.

### 4.2 Application Lifespan

We were surprised to see more than half (62%) of the 3400 applications installed by our users were uninstalled during the study. In order to understand the installation and uninstallation of applications, we define the *lifespan* of an application as the time between its installation and its uninstallation. We notice that many applications have a short lifespan, e.g., 20% uninstalled within a single day and 31% within two weeks. This shows that users often try applications and uninstall them shortly after installation.

We have found application category is a significant factor in application lifespan, as shown in Figure 3. *Games* and *social networking* exhibit a much shorter application lifespan, whereas *reference* and *news* have much longer lifespans. This can be explained by users' proclivity to both try many games and immediately decide if they like them, as well as boredom after extended playing or finishing the game. *Reference* and *news* applications exhibit much more sustained utility for users, and are inherently less prone to removal due to the dynamic content or functionality they provide.

We had expected paid applications to exhibit a much longer lifespan and lower uninstallation rate compared to their free counterparts. However, as shown in Figure 3, we were surprised that, proportionally, slightly more paid applications were uninstalled. The large number of paid application with one day lifespan shows that users frequently purchase applications which they quickly determine they dislike, losing money in the process. The larger number of paid application uninstalls in the next months can be attributed to the large number of paid games (Figure 1), which have shorter lifespans (Figure 3).

### 4.3 Usage Dynamics

Not surprisingly, we observed a significant variation between application usage amount and frequency among different users. The significant differences between users, even among the second and third quartiles, highlight the fact that the average or median user alone is unable to serve as a benchmark for mobile usage. Instead, it is necessary to consider a wide variation of users and usage.

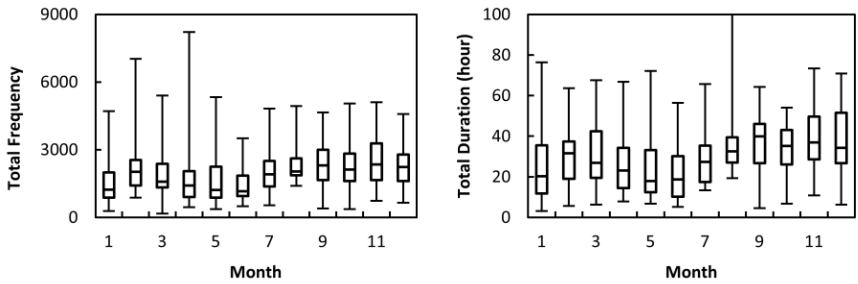
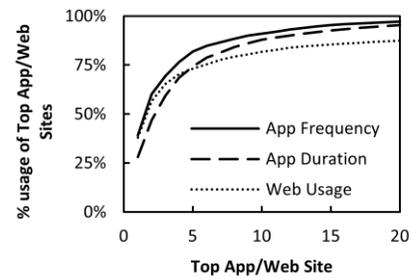

**Figure 4: Application usage is both very diverse, and increases over time, in terms of both frequency (left) and duration (right). Boxes: 2nd / 3rd quartiles. Whiskers: maximum / minimum. Horizontal lines: median**

**Figure 5: A small number of applications and websites dominate both application and web usage**

We observed a significant increase in application usage over time. Figure 4 shows the boxplot of application usage by different users, for both frequency (Left) and duration (Right). The box indicates the second and third quartiles among our users. The whiskers indicate the maximum and minimum values among our users. The horizontal lines inside the boxes indicate our users' median. We note that we did not observe a significant change in session length throughout the study. In contrast, a study in 2007-2008 using Windows Mobile smartphones highlighted an initial excitement period followed by reduced usage[5]. We attribute this difference to the much larger variety of popular 3rd party applications available on the iPhone platform at the time of the study.

Our data reveals a significant *seasonal variation* in application usage. As seen in Figure 4, application usage is significantly reduced during the summer break, coinciding months 4 through 6 (May – July). Therefore, user studies must take into account seasonal factors that affect usage.

Finally, we observe a trend for users to migrate from web based services to the iPhone application version of that service. We note that many web based services, such as Facebook, Twitter, Ebay, and Yelp, have corresponding iPhone applications. In these cases, we observed that users gradually install and use the corresponding iPhone application instead of visiting their website. For example, the median number of visits to facebook web pages decreased fourfold from the first month to the third, while the facebook application usage doubled. This finding has an important implication in the promotion of third-party applications as we will elaborate in Section 8.

### 4.4 Application Diversity

The usage of each user's top applications is a useful indicator for how diverse the application usage is. We identify each user's top applications on a monthly basis in terms of usage time and frequency. We have observed that over the course of the study, an increasing majority of usage was accounted for by the top applications. Figure 5 shows, the median percentage of usage by each user's monthly top applications. We can see that a small number of applications constitute a large share of our participant's usage in terms of frequency and duration. Approximately 40% of application usage is for the top application, and more than 90% is associated with the top 10 applications.

More importantly, we observe that diversity in application usage drops throughout the 12-month study. Figure 6 shows the median usage for applications on each user's monthly top-10 lists. It shows that our users increasingly used the applications in their monthly top-10 list, both in terms of time and frequency. The increasing dominance of each user's top applications highlights the importance of phone customizability, in order to simplify access to each user's unique top applications.

We note that there was moderate overlap between our participants' top applications. Among all users and all months, approximately seven (both average and median) of each users' top-10 applications were among the all-users-combined top-10, list for both frequency and duration.

We have found that users retain the same month-to-month similarity in application usage throughout the study, even in the first months, and despite spending an increasing portion of their time on their top applications. We calculated the similarity index between the consecutive months of each participant, shown in Figure 7. Recall from Section 3.4.1 that the similarity index is the Cosine of two usage vectors, and each usage vector is constituted of elements corresponding to the usage of each application. The median similarity remained relatively stable during the study. Interestingly, the similarity index between the first month and each month thereafter remains stable as well.

### 4.5 User Perception vs. Actual Usage

Application usage patterns tell only part of the story. Our interviews provided complementary insights into what applications the users consider as the most important components of their iPhones, and the context in which applications are used. Findings reported below have important implications for user studies of smartphones.

We have found that many of the applications used most often were not perceived as important. On the other hand, some of the least used applications were deemed extremely important by users. We asked users what applications were most important to them over the course of the entire study period. For instance, 75% of users reported the most important apps were those that helped them with their studies and functioning at school, namely the Alarm Clock, Email, the university's iPhone app, and Calendar. Games were rarely mentioned as important, even though they were frequently used.

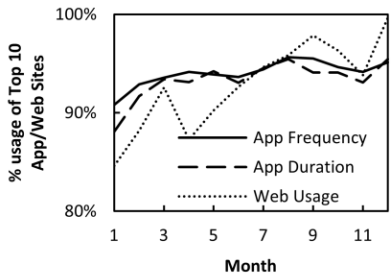
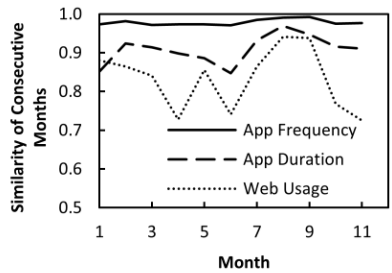
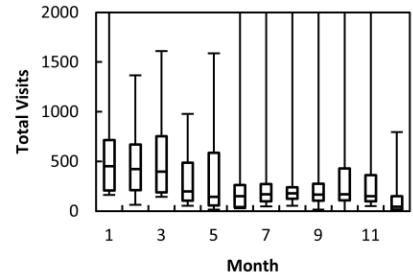

**Figure 6: Through the study, diversity in app and web usage decreases. The top 10 apps/ websites contributed to a larger fraction of median usage.**

**Figure 7: Median similarity between two consecutive months (frequency and duration for applications, frequency for websites)**

**Figure 8: Web usage is both very diverse and decreases over time. Boxplot of website visits by different users**

The interviews also gave us insight in to why users utilize particular applications. For instance, our participants utilized email largely for professional, school-related purposes. In contrast, Facebook was mostly used for personal communication. This distinction cannot be captured with logging software.

While our interviews provide additional usage information, it is important to remember the limitations of self-reporting, e.g., [14]. One such limitation is users' blind spots in reporting. For example, even though the SMS application accounted for 28% of application launches and 14% of usage duration, nobody listed this application among their most important (even though the survey specifically mentioned to consider every application currently on their iPhones). Indeed, the use of loggers such as the one employed here is important to ensure a more holistic assessment of user behaviour.

## 5. Web Usage and Dynamics

While iPhone applications are developed for a smartphone environment, and are often tailored to the specific features of the smartphone platform, we expect web browsing to be an extension and supplement to users' regular browsing since our participants had unfettered PC based web access prior to and during our study. In light of this, our data strongly suggest users are disappointed with their web browsing experience.

We have observed that several characteristics of web usage were similar to application usage. First, each user's usage converges to a small set of websites. As shown in Figure 5, the top website of a user accounts for 28% of web usage (median); and the user's top 10 websites accounts for 87%. Second, web usage diversity decreases over time, as shown by the monthly trend of the dominance of their top 10 websites, in Figure 6.

There was considerably less overlap between different users' monthly top-10 websites compared to their top-10 applications. Among all users and all months, approximately three (both average and median) of the users' top-10 websites were shared by the all-users-combined top-10 list, compared to eight for applications. Third, there is a large variation in usage patterns among users, as evident in the boxplots of web usage frequency in Figure 8. This highlights the importance of considering a wide variety of users and usage, not just the average or the median user.

In the rest of this section, we focus on the findings that are unique to web usage.

### 5.1 Usage Varies, Decreases Significantly

Contrary to application usage, we observed a significant decrease in participants' web usage throughout the study, as shown in Figure 8. This decrease strongly suggests that the users had an initial excitement regarding web browsing on the iPhone, but gradually lost their interest, probably because the web browsing on smartphones is significantly slower than that on PCs [15]. Because our participants had regular access to PC for web browsing, they were likely to be very disappointed by the web browsing experience on the iPhone. Our findings in the subsequent subsections further investigate and support this hypothesis.

Compared to application usage, we found that users were more inclined to explore web sites than applications, which is intuitive since visiting a new website requires much less commitment and time than installing an application. The key supporting evidence is the month-to-month similarity of web usage, which is significantly lower than that of application usage, as shown in Figure 7.

### 5.2 Web Content Characteristics

Our participants access both mobile and non-mobile websites. To identify the trend in mobile vs. non-mobile sites, we classify web pages based on URL keyword matching, e.g. URLs that "m.", "mobile.", "iphone.", etc. are classified as mobile. Some popular websites, such as google.com, use the same URL for both mobile and non-mobile versions. In those cases, we assume the mobile version was used.

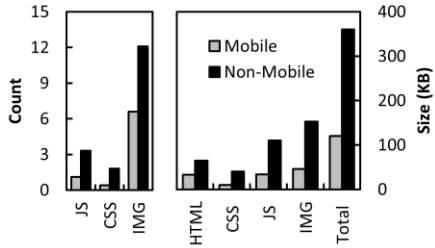
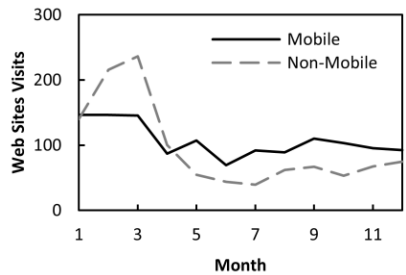
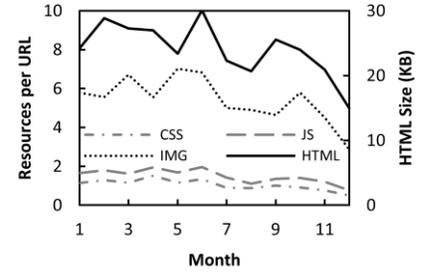

**Figure 9: Mobile web pages are less content rich, in terms of the number of resources (right) and their sizes (right)**

**Figure 10: Median visits to mobile and non-mobile website per month**

**Figure 11: Average web page resource utilization. Users prefer less content rich websites over time**

Our findings confirm that mobile web pages are less content rich than their non-mobile counterparts, in terms of styles, scripts, multimedia content, and HTML size. On average, across all users and websites, there were approximately 1.2 cascading style sheets (CSS), 1.7 JavaScript (JS) files and 9.5 images (IMG) associated with each web page. However, mobile web pages require less loading effort than non-mobile pages. As shown in Figure 9, the average HTML file size for a mobile web page was about half the size of its non-mobile. Similar to HTML files, the associated CSS, JS, and image resources of mobile web pages were also significantly fewer and smaller than their non-mobile counterparts. Overall, the smartphone had to download 120KB for the typical mobile web page and 3 times more, or 360KB, for the non-mobile web page.

We have found that over time, users prefer to visit mobile and less content-rich websites, presumably better fit for mobile devices. This is another strong indication that users are disappointed by the web browsing experience on iPhone. Our results in Figure 10 show that users initially visited more non-mobile pages, but eventually used mobile ones more often. To identify the trend in content-richness of websites, we measure the number of JavaScripts (JS), cascading style sheets (CSS), and images (IMG) in each web page, along with the size of its HTML file. To eliminate the effect of website design and layout changes over time, we analyzed all of the websites at the end of the study, and in a single day. The results are shown in Figure 11. Comparing the last three months with the first three, we can see that the median HTML size and the number of CSS, JS, and images decreased by 14%, 33%, 28%, and 19% respectively.

## 6. Socioeconomic Status (SES)

We designed the field study to study the influence of socioeconomic status (SES) on usage patterns. Yet we had not expected to see significant differences between the two SES groups of the 24 private university students, who lived in dormitories on campus and had no significant bias in their gender, major, PC access, or game console ownership. We had expected the only difference would be in how much they spent in App Store purchases. Surprisingly, our findings were the opposite of our expectations.

### 6.1 Application Usage

Application usage was consistently higher in our low SES users, approximately 40% more than high SES users in terms of both frequency and duration, as shown in Figure 12. The low SES users also consistently used a more diverse set of applications throughout the study, as shown in Figure 13 by the top 10 applications' smaller fraction of usage. The diversity is in part due to the low SES participants' higher game usage, considering the variety of games. Overall, the higher device usage and application variety in low SES users suggests that the iPhone provides more entertainment and value to the low SES users. We hypothesize that this may be due to the low SES users having fewer and less interesting *outside options*, including those for entertainment or otherwise.

There are significant differences in the applications between the two SES groups used as well. Figure 14 is a radar chart showing application usage for each of the SES groups for the top 10 applications or application categories, normalized to the overall average usage of each application. A radar chart is a convenient method of displaying multidimensional data on a two dimensional chart, where each axes represents one variable. Four applications display significant differences between the SES groups; Facebook, phone, games, and utilities.

Logistic regression confirms the significance of our findings. In the regression, we use all the monthly data from the 24 users, or 288 data points. In the first iteration, we use each of the four metrics, i.e., FRE (total frequency), PFRE (percentage frequency of monthly top 10 applications), DUR (total duration), and PDUR (percentage duration of monthly top 10 applications), as a single predictor to do regression. The results show that the coefficients of FRE and DUR are positive and the ones of PFRE and PDUR are negative. These results confirm that Low SES users are likely to have high frequency/duration of application usage and low percentage frequency/duration of top 10 application usage, which confirms our findings presented above. The p-values of all the predictors are less than 0.05, indicating that the predictors are more than 95% likely to be sig-nificant. Moreover, we compare the standardized logistic regression coefficients of each application or application category, as suggested in [16] to find out which applications are dominant predictors of SES. The results show that the top 3 dominant applications in frequency are utilities, games and phone; and top 3 in duration are Facebook, games and utilities, which comprise the exact same four applications as we observed in Figure 14.

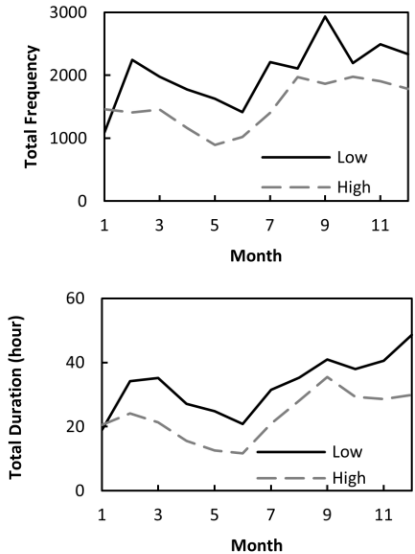 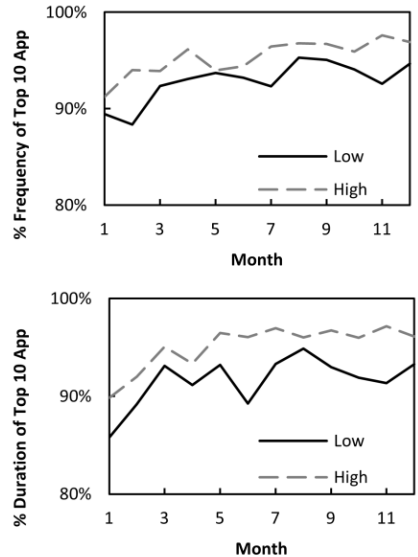 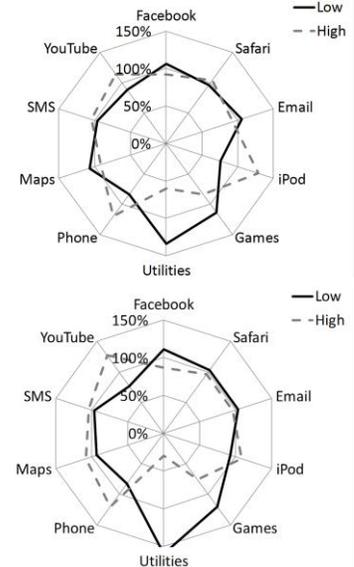

**Figure 12: Median application usage was higher for low SES participants, in terms of both frequency (top) and duration (bottom)**

**Figure 13: The top 10 applications contributed to a larger fraction of usage for high SES groups, in terms of both frequency (top) and duration (bottom)**

**Figure 14: Application usage, relative to each application's average usage, for both SES groups in terms of frequency (top) and duration (bottom)**

### 6.2 App Store Purchases

We had expected high SES participants to spend more on paid applications, but found the opposite. Low SES users spent a median of $31 on 17 applications, compared to $15 on 6 applications for the high SES users. In other words, they spent approximately twice as much money on three times as many applications.

However, we found that low SES users were more money conscious and presumably more careful in their purchases compared to high SES users. This is shown by their significantly different prices paid per hour usage of paid applications. By dividing the total each user spent in the App Store by the total paid application usage duration, we calculate the cost per hour for paid applications (price / duration). We found that low SES users had significantly lower prices paid per hour (median: $1.0 vs. $2.6), which is significant even considering the increased overall usage of the low SES users.

### 6.3 Web Usage

Web usage was initially higher in our low SES groups, showing that they had higher excitement regarding the value of mobile web access for them. However, the usage of both groups dropped, and their differences disappeared through the course of the study, as shown in Figure 15 (top). We attribute this to the shortcomings of the smartphone browsers.

In contrast with application usage, both SES groups had similar diversity in website usage, as was shown in Figure 15 (bottom). We attribute the similarity to our participants previously established web browsing habits. Both groups knew in advance which websites they wanted to visit, and those websites didn't change much.

## 7. Similarity in Usage

In this section, we use *clustering* to classify users only based on their usage pattern. The objective of our cluster analysis is to identify groups of users with similar usage patterns, and indicative applications that define their similarities and differences. Since we are not comparing the two balanced SES groups nor are we studying long-term trends, we are able to utilize our entire dataset of main and community college users for clustering. The community college participants are marked as very low SES.

### 7.1 Clustering Methodology

We use the Ward's clustering method [17] to analyze the usage patterns of all of our users. Ward's is an agglomerative hierarchical clustering method, where each user starts out as a single cluster, and the clusters are progressively merged to form larger clusters. At each stage, two clusters are chosen to be merged so that the error sum of squares, E, of the resulting cluster is minimized. E is defined as the sum of the squared distances of users from the centre of gravity of the cluster they belong to. E is initially zero, since every user is in a cluster of their own. We chose Ward's clustering method for two reasons. First, it applies strict and efficient clustering rules. While there is a large amount of overlap between users, hierarchical clustering parsimoniously separates users into distinct clusters based on representative differences in app usage. Second, it uses an iterative and efficient approach to separate users into clusters.

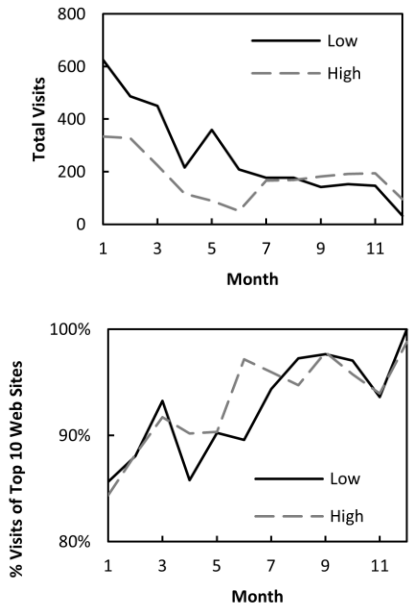 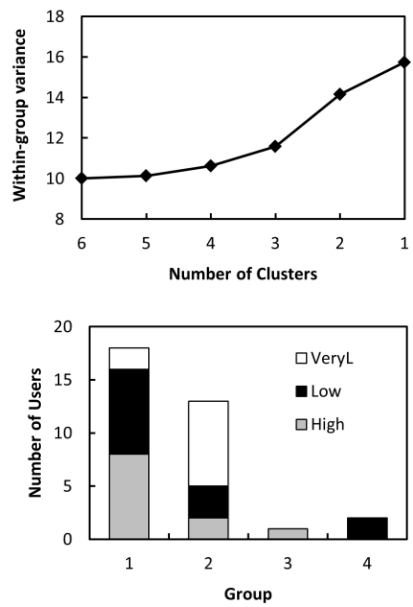 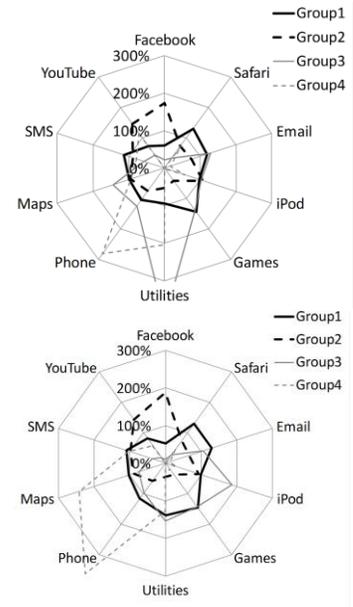

Figure 15: Median web usage was initially higher for low SES users, but became similar to high SES users (top). The top 10 websites contributed to a similar fraction of usage for both SES groups (bottom)

Figure 16: Top: The within-group variance jumps when clusters were reduced from four to three, therefore we choose four clusters. Bottom: The composition of each cluster is shown in terms of SES groups

Figure 17: Application use, relative to each application's average, for each of the four clusters in terms of frequency (top) and duration (bottom)

### 7.2 Usage Clusters

We formed a proximity matrix for each user using their normalized frequency and duration usage vectors, corresponding to the proportions of usage, along with two other indexes, their total usage duration and frequency. We eliminate the magnitude of usage, as our experiments showed that without normalization, clustering algorithms would result in clusters mainly based on the user's usage magnitude differences, and not their type of usage. In other words, the magnitude of usage would dominate and mask the comparatively subtle differences in directions of usage.

In order to determine the number of clusters, we evaluated the within-group variance at each stage of the clustering. The smaller value of this within-group index reveals how similar users are in each cluster. When disparate clusters are formed, this within-group variance jumps substantially. As can be seen in Figure 16 (top), the within-group variance begins to increase dramatically after four groups formed three groups. Thus, we concluded a four-group solution was most appropriate for our data. Two of these clusters were outliers. The breakdown of the users in each cluster is shown in Figure 16 (bottom). A large majority of Group 1 consists of the high and low SES levels (i.e., core participants. The users from the very low SES community college group mostly ended up in Group 2.

### 7.3 Identifying Representative Usage

Our clusters show significant differences in usage patterns. Figure 17 shows the application use, in comparison to the average usage of each application for the four clusters, in terms of both usage frequency and duration. It includes the top 10 applications or application categories. We can see that each cluster had their unique usage among the presented applications. In Section 10, we will present how these findings affect the design of mobile phones.

We note that among the applications presented above, four were both very popular and distinctive. These applications consumed more launches and time for some groups compared to other groups. They are the phone, iPod application, Facebook, and Safari. Thus, capturing the usage of these applications can give indication of the user's cluster.

## 8. Design Implications

A natural question to ask of a usage study paper is: *so what*? Toward answering this question, we next elaborate the implications of our findings for the design of smartphones, their applications, and evaluation studies.

### 8.1 Field Evaluation

Our study provides important insights into how the field evaluation of smartphone and its service should be designed and carried out. First, our results demonstrate the importance of carefully selecting participants in order to reveal the impact of demographic factors. Prior work on smartphone usage was not particularly prudent in participant selection and, not surprisingly, failed to reveal any difference [4], or failed to provide conclusive evidence for speculated differences [5].

Second, our results demonstrated that extraordinary care must be taken in drawing conclusions from data collected by giving out devices and studying them in field for a short period of time (e.g. shorter than three months). Our results show that the first months see a significantly different degree of exploration and diversity in usage than in the remaining months (e.g. Figures 2, 4, 6, 8, 9, and 10). Moreover, because usage continues to evolve even one year into the study, conclusions drawn using data collected from a short period of time should be generalized with care. Examples include the observed seasonal variation in usage, and popular applications losing their appeal, as is often the case with games.

Third, our study demonstrated the value of following the same users for a long period of time and of being able to interview them for insights into their behaviour. This is shown both by the significant usage changes in the later months of the study, and the discrepancy sometimes observed between self-reported and logger recorded data. However, this method is expensive financially and administratively and, therefore, can only be applied to a relatively small number of participants. As a result, this method is complementary to those that gather data from a large number of users but only sporadically, e.g., [1].

### 8.2 Application Development

Our results also provide insights into promoting third-party smartphone applications. First, our results show that smartphone users are more comfortable exploring websites and web applications than downloaded applications, as highlighted in Section 4.3. The lower month-to-month similarity in website visits compared to application usage also demonstrates the users' proclivity to explore a diverse set of websites. It is natural for users to be more adventurous in accessing different web sites than using applications; visiting a web site takes much less commitment than installing an application. This suggests that an application provider could reach a larger audience by providing a web service similar to its installation-based application when appropriate, so that first-time users can assess the application without installation.

Second, our findings regarding the application lifespan (Figure 7) show that users often try out applications for short periods, e.g. a day. Unfortunately, neither the Apple app Store nor the Android Market offers try-before-you-buy as a universal feature. Instead, users are typically expected to purchase applications based on reviews and word of mouth. However, our findings clearly indicate that users would benefit from a try-before-you-buy feature, such as the one introduced by the recent Windows Phone 7 platform. This would enable them to waste less money, as well as potentially explore and purchase more applications. Additionally, real estate on iPhones is important and a try-before-you-buy store can facilitate users to quickly "clean house" if an application isn't useful or engaging.

### 8.3 Smartphones for IT Access

Many have envisioned feature-rich smartphones that provide cost-effective access to information technologies and entertainment, especially for users from underserved communities. This was one of the key motivations for our study to focus on socioeconomic status (SES). Our results do support this vision: users with low SES tend to use smartphones more frequently and for more time than high SES users (>40% more). Our findings regarding web usage further suggest that low SES users tend to use their smartphones more often for PC roles while higher SES users tend to use their smartphones as supplemental to PCs. Preliminary findings from our community college (very low SES) participants further support this case.

On the other hand, our results also suggest, not surprisingly, that smartphones still need improvement in order to deliver satisfying performance for holistic IT access. This is evident from Figure 15, showing that users, in particular low SES ones, started with significantly higher web usage but ended with lower usage in the second half year of the study. In contrast, application usage by low SES users remained consistently higher than that by high SES users. This indicates poor experience with the web browser discouraged users, and in particular low SES users, from using the browser, and highlights the importance of improving smartphone browsers.

### 8.4 Smartphone Design

Based on the results of our clustering, we identify several key groups of users that phones must cater to. We acknowledge that we observe these clusters from a very narrow demography of smartphone users (college students), and that a broader user population likely has many more and different groups. Nonetheless, the significant differences in our narrow demographic strongly suggest that the one-size-fits-all paradigm, taken by Apple, does not serve the best interest of users. Instead, multiple mobile platforms with appropriately selected features are more likely to compliment the needs of different user groups. Hardware and/or OS vendors may achieve this through different hardware and/or OS designs, or in part through the customization of the software.

Group 1 users' higher utilization of email and utilities indicates they will enjoy features traditional to 'business' smartphones, in particular enhanced email services and a hardware keyboard. However, traditional business smartphones fail to satisfy this group, since they have a significantly higher utilization of games and the browser as well. This shows that business phones need to be equipped with a high quality browser and provide a variety of appealing games to be successful towards these users.

Group 2 users mainly use Facebook for social networking and YouTube for consuming and sharing video content. On the other hand, they use the browser, email, games, utilities, and even voice calls more sparsely. Such users will best benefit from a *social networking phone*. We have already observed attempts to develop social networking phones, to varying degrees of success, e.g. from HTC (hardware Facebook button), Motorola (customized Android builds), and Microsoft (in Windows Phone 7, now defunct Kin Phone). Our findings indicate the failure of the Microsoft Kin Phone was not due to the fact that a social networking phone lacks appeal, but due to the poor realization of its features, in particular regarding social networking itself.

Group 3 and Group 4 only have a few members. Nonetheless, and in light of our small sample size, i.e., 34, we speculate that the number of users belonging to Groups 3 and 4 is commercially significant in the general user population. Group 3 suggests the appeal of a gaming phone with media playing capabilities, such as the Sony PSP Go or Xperia Play. However, the relatively short lifespan of games, combined with the diversity number of games our participants downloaded shows that in order for a gaming phone to be successful, it needs to provide a wide range of games. The one user in Group 4 reminds of the appeal of *featurephones*, standard phones with one or a few advanced features (e.g. navigation). Many such phones exist (e.g. Garmin Nuvifone), but we hypothesize that there may be many such users but with different application needs. They would benefit from a smaller smartphone, such as the Sony Xperia X10 or the rumoured iPhone Nano.

## 9. Conclusion

We presented the findings from studying 34 iPhone 3GS users in the field. Our findings showed that the iPhone users changed over the year, highlighted the influence of socioeconomic status on device usage, and revealed how some users are similar to each other than others. Our findings have interesting implications for the design and evaluation of smartphone systems and applications, as summarized in Section 8.

Our findings probably raise more questions than they can answer in this paper. One of our objectives in writing this paper is to raise such questions to the research community. We hope that our findings will motivate researchers from multiple disciplines to work together toward answering them and, as a result, to offer even more insights into a better and more useful smartphone.